\documentstyle[pra,aps,twocolumn,psfrag,graphicx,subeqn]{revtex}

\newcommand{\Boltz}{ k_{\rm\scriptscriptstyle B}}

\newcommand{\sq}{{\sqrt{\rho}}}

\newcommand{\cov}[2]{{\langle\Delta #1\Delta #2\rangle}}

\newcommand{\p}{{\bf p}}
\newcommand{\ydot}{{\bf \dot y}}
\newcommand{\y}{{\bf y}}
\newcommand{\be}{{\bf e}}

\newcommand{\bn}{{\bf n}}
\newcommand{\egrad}{{{\bf e}'}}
\newcommand{\ugrad}{{{\bf u}'}}
\newcommand{\sgrad}{{{\bf s}'}}
\newcommand{\ort}{{{\bot \bf ye}'}}
\def\bbold#1{\mathchoice{\mbox{\boldmath $#1$}}{\mbox{\boldmath $#1$}}{\mbox{\boldmath $\scriptstyle #1$}}{\mbox{\boldmath $\scriptscriptstyle #1$}}}
\def\d{{\bbold\delta}}
\newcommand{\se}{{\rm se}}
\newcommand{\pe}{{\rm pe}}

\begin{document}

\title{A nonlinear model dynamics for closed-system, constrained,\\ maximal-entropy-generation relaxation by energy redistribution}
\author{ Gian Paolo Beretta }
\address{Universit\`a di Brescia, via Branze 38, 25123 Brescia, Italy\\
{\tt beretta@unibs.it}}
\date{\today}

\maketitle

\begin{abstract}
We discuss a nonlinear model  for the relaxation by energy
redistribution within an isolated, closed system composed of
non-interacting identical particles with energy levels $e_i$ with
$i=1,2,\dots,N$. The time-dependent occupation probabilities
$p_i(t)$ are assumed to obey the nonlinear rate equations $\tau
\,dp_i/dt=-p_i\ln p_i-\alpha (t)\,p_i-\beta(t)\,e_ip_i$ where
$\alpha(t)$ and $\beta(t)$ are functionals of the $p_i(t)$'s that
maintain invariant the mean energy $E=\sum_{i=1}^N e_i\,p_i(t)$
and the normalization condition $1=\sum_{i=1}^N p_i(t)$. The
entropy $S(t)=-\Boltz\sum_{i=1}^N p_i(t) \ln p_i(t)$ is a
non-decreasing function of time until the initially nonzero
occupation probabilities reach a Boltzmann-like canonical
distribution over the occupied energy eigenstates. Initially zero
occupation probabilities, instead, remain zero at all times. The
solutions $p_i(t)$ of the rate equations are unique and
well-defined for arbitrary initial conditions $p_i(0)$ and for all
times. Existence and uniqueness both forward and backward in time
allows the reconstruction of the ancestral or primordial lowest
entropy state. By casting the rate equations not in terms of the
$p_i$'s but of their positive square roots $\sqrt{p_i}$, they
unfold from the assumption that time evolution is at all times
along the local direction of steepest entropy ascent or,
equivalently, of maximal entropy generation. These rate equations
have the same mathematical structure and basic features of the
nonlinear dynamical equation proposed in a series of papers ended
with G.\,P.\ Beretta, Found.\ Phys.\ {\bf 17}, 365 (1987) and
recently rediscovered in S.\ Gheorghiu-Svirschevski, Phys.\ Rev.\
A {\bf 63}, 022105 and  054102 (2001).  Numerical results
illustrate the features of the dynamics and the differences with
the rate equations recently considered for the same problem in M.\
Lemanska and Z.\ Jaeger, Physica D {\bf 170}, 72 (2002). We also
interpret the functionals $\Boltz \alpha (t) $ and
$\Boltz\beta(t)$ as nonequilibrium generalizations of the
thermodynamic-equilibrium Massieu characteristic function and
inverse temperature, respectively.
\end{abstract}

\pacs{03.65.Ta,11.10.Lm,04.60.-m,05.45.-a}


\section{Introduction}

Much work has appeared in recent years  on the study of
entropy-generating irreversible nonequilibrium dynamics. Limited
discussions of previous work is found in
\cite{Lemanska,Gheorghiu2,nonlinearity} and references therein,
but no thorough critical review of the subject is available,
although it would be very helpful to provide proper
acknowledgement of pioneering work, avoid 'rediscoveries' such as
in \cite{Gheorghiu1} and outline the different frameworks,
motivations, approaches and controversial aspects. To be sure,
recent discussions \cite{Gheorghiu2,Gheorghiu1,Domokos,Czachor} on
possible fundamental tests of standard unitary quantum mechanics,
related to the existence of `spontaneous decoherence' at the
microscopic level, and on understanding and predicting decoherence
in important future applications \cite{decoherence} involving
nanometric devices, fast switching times, clock synchronization,
superdense coding, quantum computation, teleportation, quantum
cryptography, etc. show that the subject of irreversible
nonequilibrium dynamics is by no means settled.

It is not the purpose of this paper  to attempt such a difficult
review, nor to address the related fundamental issues lurking
beneath interpretation (see, e.g.,
\cite{occupation,preparation,H&G}). Rather we wish to address the
model problem recently outlined in \cite{Lemanska}.

This model may prove useful to complement various historical and
contemporary efforts to extend linear Markovian theories of
dissipative phenomena and relaxation based on master equations,
Lindblad and Langevin equations, to the nonlinear and far
nonequilibrium domain. For example, spectroscopic studies of the
effects of vibrational relaxation on line shapes of two-level
electronic transitions cannot be regularized under the Markovian
approximations  so that various nonlinear approaches are being
developed and tested \cite{Kenkre} in some cases at the expense of
giving up preservation of (complete) positivity \cite{Strunz} or
hermiticity \cite{Breuer} of the (reduced) density operator.

Again, it is not our purpose here to review the literature of
these specific potential applications of our model dynamics, nor
to apply it explicitly to particular examples. Rather we wish to
focus on illustrating its general features (including preservation
of positivity and hermiticity at all times, even backwards) that
make it a good candidate (that is, compatible with all reasonable
requirements imposed by thermodynamic principles \cite{MPLA}) of
extensions of the traditional linear master equations for open
system dynamics, capable to include the description of nonlinear
spontaneous relaxation within the system (even if isolated) by
energy redistribution between the occupied levels.

We consider an isolated,  closed system composed of
non-interacting identical particles with single-particle energy
levels $e_i$ with $i=1,2,\dots,N$ where $N$ is assumed finite for
simplicity and the $e_i$'s are repeated in case of degeneracy. We
restrict our attention on the class of dilute-Boltzmann-gas states
in which the particles are independently distributed among the $N$
(possibly degenerate) one-particle energy eigenstates. In density
operator language, this is tantamount to restricting the attention
on the subset of one-particle density operators that are diagonal
in the representation which diagonalizes the one-particle
Hamiltonian operator. We denote by $p_i$ the occupation
probability of the $i$-th eigenstate, so that the per-particle
mean energy, normalization and entropy functionals are given by
the relations
\begin{eqnarray}\label{energy}
 E(\p)&=&\sum_{i=1}^N e_ip_i\ , \qquad U(\p)=\sum_{i=1}^N p_i \ ,\nonumber\\ S(\p)&=&-\Boltz\sum_{i=1}^N p_i \ln p_i\ ,
\end{eqnarray}
where $\p$ denotes the  vector of $p_i$'s, the Boltzmann constant
$\Boltz$ may be used to nondimensionalize $S$ (or we may assume
for simplicity $\Boltz=1$ unit of entropy), and of course
$U(\p)=1$ for any normalized distribution $\p$.

As is well known, for a given value of $E$, the
thermodynamic-equilibrium canonical distribution
\begin{equation}\label{canonical}
p_j^\se (E) =\frac{\exp(-\beta^\se  (E)\,e_j)}{\sum_{i=1}^N \exp(-\beta^\se  (E)\,e_i) }
\end{equation}
has inverse temperature $\beta^\se  (E)=1/\Boltz T(E)$  and
maximal entropy $S^\se  (E)=-\Boltz\sum_{i=1}^N p_i^\se  (E) \ln
p_i^\se  (E)$.

We are interested in studying the dynamics of  a nonequilibrium
distribution obtained, for example, by exciting some energy
eigenstates. As suggested in \cite{Lemanska}, a way to alter the
distribution is to repopulate (e.g., by selective laser heating)
or depopulate (in principle, by selective cooling or resonance
fluorescence) a subset of eigenstates. This is described by
multiplying each $p_j$ by a perturbation factor $f_j\ge 0$ (with
$j=1,2,\dots,N$) (repopulation $f_j> 1$, depopulation $f_j< 1$)
and then renormalizing, to yield the perturbed nonequilibrium
distribution
\begin{equation}\label{renormalization}
\widetilde p_j =\frac{f_j\, p_j^\se  (E)}{\sum_{i=1}^N f_i\, p_i^\se  (E) }\ .
\end{equation}
Of course, in general the perturbed  distribution has a different
mean energy $\widetilde E=\sum_{i=1}^N e_i\widetilde p_i $ and
different entropy $\widetilde S=-\Boltz\sum_{i=1}^N \widetilde p_i
\ln \widetilde p_i$. However, a proper choice of the perturbation
factors $f_i$ may maintain $\widetilde E=E$, in which case
$\widetilde S< S^\se  (E)$ (see Section \ref{numerical}).

To describe the relaxation towards the  new target canonical
equilibrium distribution $ \p^\se  (\widetilde E)$, the dynamical
equation proposed in \cite{Lemanska} is, for $j=1,2,\dots,N $,
\begin{subequations}\label{Lemanska1}
\begin{equation}
\frac{dp_j}{dt}=-\upsilon\left[\ln p_j+a_L(\p) +b_L(\p)\,e_j \right] \ ,\end{equation} where
\begin{eqnarray} a_L(\p)&=& \frac{ \sum_i e_i\, \sum_j e_j \ln p_j- \sum_i \ln p_i\, \sum_j e_j^2 }{ N \sum_i e_i^2  - \Big(\sum_i e_i \Big)^2 } \ ,\\
 b_L(\p)&=& \frac{ \sum_i \ln p_i\, \sum_j e_j - \sum_i e_i \ln p_i}{N  \sum_i e_i^2  -\Big( \sum_i e_i \Big)^2 } \ .
\end{eqnarray}
\end{subequations}
This equation does have the capability of  continuously
rearranging the distribution so that the perturbed distribution
evolves towards the maximal entropy target distribution given by
Eq.\ (\ref{canonical}) with energy $\widetilde E$. However, in the
far-nonequilibrium region it has the defect to imply the
unphysical feature that an initially unpopulated eigenstate gets
populated at an infinite rate. This feature is in contrast with a
wealth of successful models of physical systems in which by
limiting our attention to a subset of relevant single-particle
eigenstates we get good results, that are relatively robust with
respect to adding to the model other less relevant, unpopulated or
little populated eigenstates. According to Eq.\ \ref{Lemanska1},
instead, distributions where some eigenstates are very little
populated would survive only for extremely short times.

The equation of motion that we propose for  the time evolution of
the perturbed distribution is, for $ j=1,2,\dots,N $,
\begin{subequations}\label{mainEquation1}
\begin{equation}
\frac{dp_j}{dt}=-\frac{1}{\tau}\left[p_j\ln p_j+\alpha(\p)\,p_j+\beta (\p)\,e_jp_j\right] \ ,\end{equation} where
\begin{eqnarray} \alpha (\p)&=& \frac{\sum_i e_ip_i\,\sum_j e_jp_j\ln p_j- \sum_i p_i\ln p_i\,\sum_j e_j^2p_j}{ \sum_i e_i^2p_i -\Big(\sum_i e_ip_i\Big)^2 }\ , \\
 \beta(\p)&=& \frac{\sum_i p_i\ln p_i\,\sum_j e_jp_j-\sum_i e_ip_i\ln p_i}{ \sum_i e_i^2p_i -\Big(\sum_i e_ip_i\Big)^2 } \ .
\end{eqnarray}
\end{subequations}

We show in Section \ref{Equation} that the  apparently slight
modification with respect to Eq.\ (\ref{Lemanska1}) not only fixes
the cited defect, while maintaining the relevant overall features
of conserving energy, normalization, nonnegativity of the
probabilities and maintaining the entropy generation rate
nonnegative. It also features existence and uniqueness of the
solutions of the Cauchy problem for all times, $-\infty<t<+\infty
$, and entails a large class of partially-canonical equilibrium
distributions that are unstable, as well as a single
conditionally-stable canonical equilibrium distribution for each
value of the energy, as required by a well-known statement of the
second law of thermodynamics \cite{Books,Lyapunov}.

We show in Section \ref{Equation} that the  structure of Eq.\
(\ref{mainEquation1}) is the same as that of the general nonlinear
quantum equation we discuss in a series of papers written over
twenty years ago
\cite{Cimento1,Cimento2,Beretta,Attractor,Fluorescence} in which
we develop and propose a nonlinear quantum dynamics in an attempt
to unite ordinary quantum mechanics and general equilibrium and
nonequilibrium thermodynamics. As acknowledged also in
\cite{Gheorghiu2,Korsch}, the nonlinear quantum dynamical law
first proposed by this author \cite{thesis} does have very
intriguing and appealing mathematical features. We must admit
however that the physical interpretation, motivation and context
of our pioneering scheme is still considered 'adventurous'
\cite{Nature} by most of the physical community, although we would
prefer to term it 'revolutionary' in the sense of Kuhn
\cite{Kuhn}. For this reason, in this paper we do not pursue such
controversial interpretation, but we wish to emphasize that ---
leaving aside its interpretation and focusing attention only on
its mathematics --- our previous work represents to our knowledge
the first time that the steepest-entropy-ascent (or
maximal-entropy-generation) ansatz has been explicitly formulated
and implemented in a general dynamical law capable of describing
the relaxation of arbitrary nonequilibrium states towards
thermodynamic equilibrium.

In Section \ref{construction} we provide  a derivation of Eq.\
(\ref{mainEquation1}) from the assumption that the occupation
probability distribution evolves along the steepest-entropy-ascent
trajectory in the state space defined in terms not of the $p_i$'s
but of their positive square roots $\sqrt{p_i}$'s. In Section
\ref{fluctuation} we derive a fluctuation-dissipation formulation
of the equation and in Section \ref{variational} a variational
formulation.

In Section \ref{Twolevel} we discuss a  simplest degenerate case
in which the relaxation equation admits an analytical solution,
and we compare results with those numerically derived from the
natural extension of Eq.\ (\ref{Lemanska1}) to such case. Finally,
in Section \ref{numerical} we show some numerical results that
illustrate the general features of the proposed nonlinear
relaxation equation.

\section{Main features of the assumed nonlinear relaxation equation}\label{Equation}

By analogy with the dynamical law  introduced in
\cite{Cimento1,Beretta,thesis,Onsager}, Eq.\ (\ref{mainEquation1})
may be also written as a ratio of determinants in the form
\begin{equation}\label{mainEquation}\frac{dp_j}{dt}=-\frac{1}{\tau}\frac{\left|\begin{array}{ccccc} p_j\ln p_j &  & p_j &  & e_jp_j \bigskip \nonumber\\ {\sum p_i\ln p_i} &  & {\displaystyle 1} &  & {\sum e_ip_i } \bigskip \nonumber\\ {\sum e_ip_i\ln p_i }&  &  {\sum e_ip_i }&  & {\sum e_i^2p_i} \nonumber\end{array} \right|}{\left|\begin{array}{ccc}
{\displaystyle 1}\vphantom{\sum^A}&  & {\sum e_ip_i} \bigskip \nonumber\\
{ \sum e_ip_i }&  & {\sum e_i^2p_i }\nonumber
\end{array}
\right| } \ ,\end{equation} where $|\cdot|=\det[\cdot]$, and
$\tau$  is assumed constant \cite{tau} and may be used to
nondimensionalize time (or we may assume $\tau=1$ unit of time).

The resulting rate of entropy generation  may be written as a
ratio of Gram determinants in the form
\begin{equation}\label{entropyRate}\frac{dS}{dt}=\frac{\Boltz}{\tau}\frac{\left|\begin{array}{ccccc}  {\sum p_i(\ln p_i)^2 }& & {\sum p_i\ln p_i } & &  {\sum e_ip_i\ln p_i } \bigskip \nonumber\\ {\sum p_i\ln p_i}  & & {\displaystyle 1}  & & {\sum e_ip_i } \bigskip \nonumber\\ {\sum e_ip_i\ln p_i } & &  {\sum e_ip_i }  & & {\sum e_i^2p_i} \nonumber\end{array} \right|}{\left|\begin{array}{ccc}
{\displaystyle 1}\vphantom{\sum^A} & & {\sum e_ip_i} \bigskip \nonumber\\
{\sum e_ip_i }  & & {\sum e_i^2p_i }\nonumber
\end{array}
\right| } \ge 0\ ,\end{equation} where the non-negativity follows
from the  well-known properties of Gram determinants (see also
Section \ref{construction}).

Eq.\ (\ref{mainEquation1}) or the equivalent Eq.\ (\ref{mainEquation}) is well-behaved in the sense that the following general features  can be readily verified (detailed  proofs in \cite{Gheorghiu1,Cimento1}):
\begin{itemize}
\item it conserves the normalization of the distribution and the mean energy $E$ along the entire time evolution;
\item it preserves the non-negativity of each $p_i$;
\item it maintains the rate of entropy generation non-negative at all times;
\item it maintains unoccupied all the initially unoccupied eigenstates; in other words, given a distribution $p_i$ and defining the vector $\d(\p)$ of $\delta_i$'s such that, for each $i=1,2,\dots,N$, $\delta_i=0$ if $p_i=0$ or $\delta_i=1$ if $p_i\ne 0$, the vector $\d$ is time invariant;
\item it drives any arbitrary initial distribution $\p (0)$ towards the partially-canonical (or canonical, if $\delta_i=1$ for all $p_i$'s) equilibrium distribution, reached as $t\rightarrow\infty $,
\begin{equation}\label{nondissipative}
p_j^\pe (E,\d) =\frac{\delta_j\exp(-\beta^\pe  (E,\d)\,e_j)}{\sum_{i=1}^N \delta_i \exp(-\beta^\pe  (E,\d)\,e_i) }\ , \end{equation}
where, of course, $\d=\d (\p(0)) $ and the value of $\beta^\pe $ is determined by the initial state through the relation  $\sum_{i=1}^N e_i \,p_i^\pe (E,\d) =E=E(\p(0))$. Distributions (\ref{nondissipative}) are those for which $d\p/dt=0$, i.e., that satisfy the equilibrium condition $ p_i\ln p_i =-\alpha p_i -\beta e_ip_i$ for all'$i$'s and some scalars $\alpha$ and $\beta$.
\end{itemize}

Moreover, Eq.\ (\ref{mainEquation}) is  well-behaved not only in
forward time but also in backward time, consistently with the
strongest form of the principle of causality, by which future
states of a strictly isolated system should unfold
deterministically from initial states along smooth unique
trajectories in state domain defined for all times (future as well
as past). Indeed, for any given arbitrary 'initial' distribution
$\p(0)$ we can follow the unique trajectory $\p(t)$ for
$-\infty<t<+\infty $. In forward time the target distributions of
all trajectories are given by Eq. (\ref{nondissipative}),
$\p(+\infty)= \p^\pe (E(\p(0)),\d (\p(0)))$. The backward-time
earliest (or 'primordial') lowest-entropy distribution
$\p(-\infty)$ is also uniquely identified by the given initial
distribution $\p(0)$ through Eq.\ (\ref{mainEquation}),  but it is
harder to characterize analytically in general. Depending on the
given $\p(0)$, the redistribution among energy eigenstates may
affect some of the occupation numbers in a non-monotonic way. In
the limit as $t\rightarrow -\infty $, however, all $dp_j/dt$'s
[the rhs of each of Eqs.\ (\ref{mainEquation})] become
sign-definite; for example, they may become all positive except
for a particular one which tends to $p_{\overline k}(-\infty)=1$,
so that all others tend to zero, $p_{j\ne\overline k}(-\infty)=0$
[this can happen only if the mean energy $ E(\p(0))$ is exactly
equal to the $\overline k$-th energy level, i.e., only if
$E(\p(0))=e_{\overline k}$], or they may all tend to zero except
for two particular $p_j$'s, say $p_{\overline j}$ and
$p_{\overline k}$ which tend to finite values, clearly with
$p_{\overline j}+p_{\overline k}=1$ (examples in Section
\ref{numerical}).

Because the model equation maintains the  rate of entropy
generation non-negative, the entropy functional $S$ [Eq.\
(\ref{energy})] is an $S$-function \cite{Lyapunov} and, therefore,
every thermal-like canonical equilibrium distribution
(\ref{nondissipative}) is $\d$-$E$-conditionally stable, that is,
stable with respect to perturbations that do not alter the mean
value $E$ of the energy and the set of unoccupied energy
eigenstates (described by the zeroes in vector $\d$). These
distributions constitute the `target' highest-entropy states
compatible with the mean value of the energy and the invariant
subset of unoccupied eigenstates. These distributions, however,
are not $E$-conditionally stable, that is, stable with respect to
all perturbations that do not alter the mean value $E$. Indeed,
starting from a distribution (\ref{nondissipative}), a
perturbation that changes a zero probability to an infinitesimal
value, makes the perturbed distribution proceed in time by
amplifying that probability until a new, different and
higher-entropy, target canonical distribution is reached.  For a
given mean energy $E$, the only canonical distribution that is
$E$-conditionally stable is the one for which all energy
eigenstates are occupied, i.e., the maximal-entropy canonical
distribution (\ref{canonical}).

By interpreting the entropy $S$ as a  measure of how `well' the
energy is distributed among the available energy eigenstates, the
proposed nonlinear dynamics describes a spontaneous internal
redistribution of the energy along the path of maximal entropy
increase leading towards an `optimally' distributed
(highest-entropy) state compatible with the condition of
maintaining unoccupied the initially unoccupied energy
eigenstates.

\section{Construction of the equation from the steepest-entropy-ascent ansatz}\label{construction}

In this section, we provide a brief  derivation of Eq.\
(\ref{mainEquation}) from the assumption that the occupation
probability distribution evolves along the steepest-entropy-ascent
trajectory in the proper state space. We also discuss an important
degenerate case.

For the purpose of this derivation,  instead of working with the
vector $\p$ of the occupation probabilities, we work in terms of
their positive square roots \cite{Reznik}, $y_i=\sqrt{p_i}$ and
the corresponding vector $\y$. We rewrite the mean energy,
normalization and entropy functionals as
\begin{eqnarray}\label{energyy}
 E(\y)&=&\sum_{i=1}^N e_iy^2_i \ ,\qquad  U(\y)=\sum_{i=1}^N y^2_i \ ,\nonumber\\ S(\y)&=&-\Boltz\sum_{i=1}^N y^2_i \ln y^2_i \,
\end{eqnarray}
where, of course, $U(\y)=1$ for any $\y$.
The $i$-th component of the gradients of these functionals are, respectively,
\begin{equation}\label{gradients}
 e'_i=2\, e_iy_i \ ,\ \ u'_i=2\, y_i \ ,\ \ s'_i=-2\Boltz\,( y_i \ln y^2_i+ y_i )
\end{equation}
and, therefore, the time-rate-of-change functionals may be written as
\begin{eqnarray}\label{rates}
 \dot E&=&(\ydot,\egrad) \ ,\qquad   \dot U=2\,(\ydot,\y) \ ,\nonumber\\
\dot S&=&(\ydot,\sgrad) \ ,
\end{eqnarray}
where $(\cdot,\cdot)$ denotes the scalar  product of two vectors
[e.g., the normalization condition $U(\y)=1$ may be rewritten as
$(\y,\y)=1$], and the energy and entropy gradient vectors $\egrad$
and $\sgrad$ are defined by the components in (\ref{gradients}),
while we choose to substitute immediately the obvious relation
$\ugrad=2\y$.

In order to maintain $(\ydot,\y)=0$  and $(\ydot,\egrad)=0$ the
vector $\ydot$ must be orthogonal to the linear manifold spanned
by $\y$ and $\egrad$.

For unconstrained maximal entropy generation,  $\ydot$ would be in
the direction of the gradient $\sgrad$ of the entropy functional
$S(\y)$; in such case, however, because $\sgrad$ is almost never
orthogonal to the $\y\egrad$ manifold, in general $U(\y)$ and
$E(\y)$ would not remain time invariant. Instead, we assume
constrained --- constant $E(\y)$ and $U(\y)$ --- maximal entropy
generation. We obtain it by taking $\ydot$ in the direction of the
component of $\sgrad$ orthogonal to the $\y\egrad$ manifold.
Denoting such component by $\sgrad_\ort$ we therefore assume
\begin{equation}\label{ydot}
\ydot=\frac{1}{4\Boltz \tau(\y)}\ \sgrad_\ort \ ,
\end{equation}
where $\tau(\y)$ may be any positive  definite functional of $\y$
with dimensions of time, that determines the time rate at which
$\y$ evolves along the path of constrained steepest entropy
ascent. For simplicity, and for the purpose of comparison with
\cite{Lemanska}, we assume $\tau$ a positive constant  as done in
our first proposal of this equation of motion in
\cite{Cimento1,Attractor,Fluorescence,thesis}.

Using the well-known theory of Gram  determinants, we can write an
explicit expression for $\sgrad_\ort$. If $\y$ and $\egrad$ are
linearly independent, we have
\begin{subequations}
\begin{equation}\label{sgradort}
\sgrad_\ort=\frac{\left|\begin{array}{ccccc}\sgrad &  & \y &  & \egrad \bigskip \nonumber\\ {\displaystyle (\sgrad,\y)} &  & {\displaystyle (\y,\y)} &  & {\displaystyle (\egrad,\y) } \bigskip \nonumber\\ {\displaystyle (\sgrad,\egrad) }&  &  {\displaystyle (\y,\egrad) }&  & {\displaystyle (\egrad,\egrad)} \nonumber \end{array}\right| }{\left|
\begin{array}{ccc}
{\displaystyle (\y,\y) \vphantom{\egrad^A}}&  & {\displaystyle (\egrad,\y)} \bigskip \nonumber\\
{\displaystyle (\y,\egrad)}&  & {\displaystyle (\egrad,\egrad)}\nonumber
\end{array}
\right| } \ .
\end{equation}
If instead $\y$ and $\egrad$ are linearly  dependent, i.e., if
$\egrad=2e\,\y$ for some scalar $e$, the expression is
\begin{equation}\label{sgradortlindep}
\sgrad_\ort=\left|\begin{array}{ccc}\sgrad &  & \y  \bigskip \nonumber\\ {\displaystyle (\sgrad,\y)} &  & {\displaystyle (\y,\y)} \nonumber \end{array}\right| \left/
{\displaystyle (\y,\y) } \right. = \sgrad-(\sgrad,\y)\,\y \ ,
\end{equation}
\end{subequations}
where we use $(\y,\y)=1$. In either case, we readily verify that
\begin{eqnarray}\label{yrates}
 \dot E&=&(\ydot,\egrad)=0 \ ,\qquad  \dot U=2\,(\ydot,\y)=0 \ ,\nonumber\\ \dot S&=&(\ydot,\sgrad)=4\tau\Boltz\,(\ydot,\ydot)\ ,
\end{eqnarray}
from which we see that the rate of entropy  generation is related
to the norm of $\ydot$ and is positive definite.

Combining Eqs.\ (\ref{ydot}) and (\ref{sgradort}) we find
\begin{subequations}\label{mainEquation2}\begin{equation} 4\Boltz\tau\,\ydot=\sgrad-a(\y)\,\y-b(\y)\,\egrad \ ,\end{equation}
where
\begin{eqnarray}
a(\y)&=&\frac{ (\sgrad,\y)(\egrad,\egrad)- (\sgrad,\egrad)(\egrad,\y)}{(\y,\y)(\egrad,\egrad)-(\y,\egrad)(\egrad,\y)}\label{alpha}\ ,\\
b(\y)&=&\frac{(\sgrad,\egrad)(\y,\y)- (\sgrad,\y)(\y,\egrad)}{(\y,\y)(\egrad,\egrad)-(\y,\egrad)(\egrad,\y)}\label{beta}\ ,
\end{eqnarray}
\end{subequations}
and, setting back $p_i=y_i^2$ and  $\dot p_i=2y_i\dot y_i$ we
readily obtain Eq.\ (\ref{mainEquation1}) and the identities
$\alpha(\p)= 1+a(\y)/2\Boltz$ and $\beta(\p)=b(\y)/\Boltz$.

Similarly, combining Eqs.\ (\ref{ydot})  and
(\ref{sgradortlindep}) we find
\begin{subequations}\begin{equation}\label{mainEquationdegenerate2} 4\Boltz\tau\,\ydot=\sgrad-(\sgrad,\y)\,\y \end{equation}
and, therefore, in the degenerate case of $\y$ and $\egrad$ linearly dependent, the relaxation equations are, for $ j=1,2,\dots,N $,
\begin{equation}\label{degenerate}
\frac{dp_j}{dt}=-\frac{1}{\tau}\left[p_j\ln p_j- p_j\left({\scriptstyle\sum}_i p_i\ln p_i\right)\right] \ ,\end{equation}
or, equivalently,
\begin{equation}\label{Gramdegenerate}\frac{dp_j}{dt}=-\frac{1}{\tau}\,\left|\begin{array}{ccc}  {\displaystyle p_j\ln p_j }&  & {\displaystyle p_j } \bigskip \nonumber\\ {\sum p_i\ln p_i} &  & {\displaystyle 1} \nonumber \end{array} \right| \ .\end{equation}
\end{subequations}
In this case, the rate of entropy generation may be written as
\begin{equation}\label{entropyRatedegenerate}\frac{dS}{dt}=\frac{\Boltz}{\tau}\,\left|\begin{array}{ccc}  {\sum p_i(\ln p_i)^2 }&  & {\sum p_i\ln p_i } \bigskip \nonumber\\ {\sum p_i\ln p_i} &  & {\displaystyle 1} \nonumber \end{array} \right|
 \ge 0\ ,\end{equation}
where the nonegativity follows from the  well-known properties of
Gram determinants.

Equation (\ref{Gramdegenerate})  substitutes (\ref{mainEquation})
when $\egrad=2e\,\y$ for some scalar $e$ or, equivalently, when
$e_ip_i=e\,p_i$ for every $i$, that is, when the populated
eigenstates all correspond to the same energy level. If satisfied
at one instant in time this condition is satisfied at all times,
both forward and backward in time. It follows that in such
degenerate cases the entire time evolution is governed by Eq.\
(\ref{degenerate}). In the general nondegenerate cases, i.e., when
at one time (and, hence, at all times) $e_ip_i\ne E\,p_i$ for two
or more $i$'s, where $E$ is the mean energy, the time evolution is
entirely governed by Eq.\ (\ref{mainEquation1}) [or, equivalently,
(\ref{mainEquation}) or (\ref{mainEquation2})].  This also implies
that the denominators of $\alpha(\p)$ and $\beta(\p)$ [or,
equivalently, of $a(\y)$ and $b(\y)$] remain positive definite at
all times  and, hence, the entire time evolution is well-defined.

As stated above, the feature that  unpopulated eigenstates remain
unpopulated is extremely important as it is compatible, for
example, with the widely accepted and successful possibility to
describe real systems by means of simplified models with a limited
number of relevant energy eigenstates.

The same feature does not hold for  equation (\ref{Lemanska1}),
because it implies that an initially unpopulated eigenstate gets
populated at an infinite rate. For the same reason, Eq.\
(\ref{Lemanska1}) does not allow tracing the time evolution
backward in time beyond the instant when the first eigenstate
becomes unpopulated, for at earlier times the condition $p_i\ge 0$
is not satisfied.

\section{Fluctuation-dissipation formulation}\label{fluctuation}

It is noteworthy that Eq.\ (\ref{mainEquation})  admits a general
fluctuation-dissipation formulation and interpretation. To see
this, we introduce the energy and entropy fluctuation functionals
as follows
\begin{eqnarray}
\cov{E}{E}(\p)&=& {\scriptstyle\sum}_i\ p_i\,[e_i-E(\p)]^2\nonumber \\ &=&{\scriptstyle\sum}_i\ p_i\,e_i^2-\Big({\scriptstyle\sum}_i\ p_i\,e_i\Big)^2 \nonumber\\  &=&\frac{1}{4}\,\left|
\begin{array}{ccc}
{\displaystyle (\y,\y) }&  & {\displaystyle (\egrad,\y)} \bigskip \\
{\displaystyle (\y,\egrad)}&  & {\displaystyle (\egrad,\egrad)}
\end{array}
\right| \ ,\\
\cov{S}{S}(\p) &=&{\scriptstyle\sum}_i\ p_i\,[-\Boltz\ln p_i-S(\p)]^2\nonumber\\ &=&\Boltz^2{\scriptstyle\sum}_i\ p_i\,(\ln p_i)^2-\Boltz^2\Big({\scriptstyle\sum}_i\ p_i\,\ln p_i \Big)^2 \ ,\\
\cov{E}{S} (\p) &=& {\scriptstyle\sum}_i\ p_i\,[e_i-E(\p)] [-\Boltz\ln p_i-S(\p)] \ ,
\end{eqnarray}
and rewrite functionals $\beta(\p)$ and $\alpha(\p)$ as
\begin{eqnarray}
\beta(\p)&=&\frac{1}{\Boltz}\frac{\cov{E}{S} (\p)}{ \cov{E}{E}(\p)} \ ,\label{betagen}\\
\alpha(\p)&=&\frac{S(\p)}{\Boltz}-\beta(\p)\,E(\p) \ .
\end{eqnarray}

At the thermodynamic equilibrium distribution  with energy $E$,
$\p^\se (E)$, we have
\begin{eqnarray}
\Boltz\beta(\p^\se (E))&=&\Boltz\beta^\se (E) =1/ T(E)\ , \\
\Boltz\alpha(\p^\se (E))&=&
\Boltz \alpha^\se(E) =S^\se (E)- E/T(E) \ ,\label{alphase}
\end{eqnarray}
and in Eq.\ (\ref{alphase}) we recognize the thermodynamic-equilibrium Massieu characteristic function \cite{Books}
\begin{equation} M^\se=S-E/T \ .\end{equation}

It is therefore natural, in this framework,  to adopt the
following generalization of the Massieu function to arbitrary
nonequilibrium distributions
\begin{equation}
M(\p)= \Boltz\, \alpha(\p) =S(\p)- \Boltz \beta(\p) E(\p) \ ,
\end{equation}
with $\beta(\p)$ given by Eq.\ (\ref{betagen}).
The corresponding fluctuations functional is
\begin{equation}
\cov{M}{M}(\p)= \sum_i\ p_i\,[ -\Boltz\ln p_i
-\Boltz\beta(\p)e_i-M(\p)]^2 \ .
\end{equation}
We can readily verify that [Eq.\ (\ref{entropyRate})] may be rewritten as
\begin{equation}\label{fluctuationdissipation}
\frac{dS}{dt}=\frac{1}{\Boltz \tau}\,\cov{M}{M}
\end{equation}
and, therefore, the rate of entropy generation  is directly
proportional to the fluctuations of our generalized Massieu
function \cite{Massieu}. Such fluctuations are related to entropy
and energy fluctuations through functional $\Boltz \beta(\p)$ as
follows
\begin{equation}\label{Massieufluctuations}
\cov{M}{M} (\p)=\cov{S}{S} (\p)-\Boltz^2\beta(\p)^2 \cov{E}{E} (\p) \ ,
\end{equation}
and become zero at every canonical  thermodynamic-equilibrium
distribution $\p^\se(E)$, Eq.\ (\ref{canonical}), and at every
partially-canonical equilibrium distribution $\p^\pe(E,\d)$, Eq.\
(\ref{nondissipative}), as well.

It is noteworthy that the  functional $\Boltz\beta(\p)$, which is
well-defined by Eq.\ (\ref{betagen}) only for distributions with
$\cov{E}{E}\ne 0$, may be interpreted in this framework as a
natural generalization to nonequilibrium of the inverse
temperature, at least inasfar as for $t\rightarrow +\infty $ it
tends to the thermodynamic-equilibrium inverse temperature
$\Boltz\beta^\se$ of distribution (\ref{canonical}) or the partial
equilibrium inverse temperature $\Boltz\beta^\pe$ of distribution
(\ref{nondissipative}).

The special case of distributions  with $\cov{E}{E}=0$ happens if
and only if $\egrad=2e\y$ for some scalar $e$ (see Section
\ref{construction}). In such special degenerate case, $\cov{E}{E}$
remains zero along the entire time evolution, which is given by
Eq.\ (\ref{degenerate}), and the role of the Massieu function is
taken up by the entropy $S$, for both equilibrium and
nonequilibrium distributions. The fluctuation-dissipation form of
the rate of entropy generation [Eq.\
(\ref{entropyRatedegenerate})] becomes  the following
\begin{equation}
\frac{dS}{dt}=\frac{1}{\Boltz \tau}\,\cov{S}{S} \ ,
\end{equation}
and in this special degenerate case  the canonical and
partially-canonical equilibrium distributions all have
$\cov{S}{S}=0$, for they consist of $N_\d=(\d,\d)$ probabilities
$p_i$ all equal to $1/N_\d$ and of $N-N_\d$ all equal to zero.

As regards the fluctuation-dissipation relations, the various well
formulated arguments, derivations and interpretations discussed
for Eq.\ (\ref{Lemanska1}) by Englman in the Appendix of Ref.\
\cite{Lemanska} and based on the steepest-entropy-ascent ansatz --
first introduced in quantum thermodynamics by the present author
\cite{Beretta} -- apply with minor modifications also for our
better-behaved dynamical equation Eq.\ (\ref{mainEquation1}). In
addition, we prove in \cite{Onsager} that  Eq.\
(\ref{mainEquation1}) implies a generalized Onsager reciprocity
theorem.

\section{Variational formulation}\label{variational}

In terms of the $y_i=\sqrt{p_i}$ notation, we can derive our
equation of motion also  as a result of the following equivalent
variational formulation (along the lines recently proposed in
\cite{Gheorghiu1})
\begin{eqnarray} \label{maxprob}
\max_{\ydot}\ \dot S=(\ydot,\sgrad)\quad {\rm subject\ to\ } \dot E=(\ydot,\egrad)=0,\nonumber\\ \dot U=(\ydot,\y)=0,\ {\rm and\ } (\ydot,\ydot)=\xi(\y)\ ,
\end{eqnarray}
where the last constraint implies  that we maximize the entropy
generation rate only with respect to the `direction' of $\ydot$,
i.e., at every given $\y$ we select the maximizing $\ydot$ among a
subset of vectors that share the same (but otherwise arbitrary)
norm $\xi(\y)$. For $\y$ and $\egrad$ linearly independent, using
the standard method, we associate the Lagrange multipliers $a$,
$b$ and $4\Boltz\tau$ with the constraints, and from Eq.\
(\ref{yrates}) and the necessary Euler-Lagrange conditions
\begin{equation}
\frac{\partial }{\partial \ydot}\left[(\ydot,\sgrad)-a\,(\ydot,\y) -b\,(\ydot,\egrad) -4\Boltz\tau\,(\ydot,\ydot)\right]=0 \ ,
\end{equation}
we readily obtain Eq.\ (\ref{mainEquation2}) as  well as, upon
substitution into the constraints, the multipliers given by Eqs.\
(\ref{alpha}) and (\ref{beta}), and the square norm of $\ydot$,
\begin{eqnarray}
\xi(\y)&=&\frac{\dot S}{4\Boltz\tau}=\frac{1}{16\Boltz^2\tau^2}\frac{\Gamma(\sgrad,\y,\egrad)}{ \Gamma(\y,\egrad)} \nonumber\\ &=& \frac{1}{16\Boltz^2\tau^2}\frac{\left|\begin{array}{ccc}(\sgrad,\sgrad) & (\y,\sgrad) & (\egrad,\sgrad) \bigskip \\ {\displaystyle (\sgrad,\y)} & {\displaystyle (\y,\y)} & {\displaystyle (\egrad,\y) } \bigskip \\ {\displaystyle (\sgrad,\egrad) } &  {\displaystyle (\y,\egrad) } & {\displaystyle (\egrad,\egrad)} \end{array}\right|
}{\left|
\begin{array}{cc}
{\displaystyle (\y,\y) }\vphantom{\egrad^A} & {\displaystyle (\egrad,\y)} \bigskip \\
{\displaystyle (\y,\egrad)} & {\displaystyle (\egrad,\egrad)}
\end{array}
\right| } \ ,
\end{eqnarray}
where $\Gamma$ denotes the Gram determinant of the argument vectors.

Similarly, for the degenerate cases  with $\y$ and $\egrad$
linearly dependent, the normalization and constant energy
conditions collapse into a unique constraint with which we
associate the Lagrange multiplier $c$, and by the same standard
procedure we obtain Eq.\ (\ref{mainEquationdegenerate2}) and, upon
substitution into the constraints, the multiplier $c=(\sgrad,\y)$,
and the square norm of $\ydot$,
\begin{eqnarray}
\xi(\y)&=&\frac{\dot S}{4\Boltz\tau}=\frac{1}{16\Boltz^2\tau^2}\Gamma(\sgrad,\y)\nonumber\\  &=& \frac{1}{16\Boltz^2\tau^2}\left|\begin{array}{ccc}(\sgrad,\sgrad) & (\y,\sgrad) \bigskip \\ {\displaystyle (\sgrad,\y)} & {\displaystyle (\y,\y)} \end{array}\right| \ .
\end{eqnarray}

\section{Simplest case: two-level  particles with degenerate eigenstates}\label{Twolevel}
The simplest mathematical form of  the model equation that derives
from the equation of motion proposed in the previous sections, is
obtained when we have an isolated, closed gas composed of
non-interacting identical two-level particles with degenerate
energy levels, such as electonic spins in the absence of an
applied magnetic field. Then $N=2$, both levels have energy
$e_1=e_2=e$, the occupation probabilities of the two corresponding
eigenstates are $p_1=1-p$  and $p_2=p$, respectively, and the
model equation (\ref{Gramdegenerate}) for redistribution among the
two eigenstates  becomes
\begin{equation}\label{twoleveldegenerateEquation}
\frac{dp }{dt}=p\,(1-p)\,\ln\frac{1-p}{p}\ ,
\end{equation}
where we set $\tau=1$. The rate  of entropy generation
($\Boltz=1$) is \begin{equation}\label{entropyratetwolevel}\dot S
= p\,(1-p)\,\left(\ln\frac{1-p}{p}\right)^2\ . \end{equation}

Not only equation (\ref{twoleveldegenerateEquation})  is
well-behaved at all times (existence and uniqueness of the
solution for any initial $p(0)$ with $0\le p(0)\le 1$), but it can
also be integrated to yield
\begin{equation}\label{solution1}
t=\int_{p(0)}^{p(t)} \frac{dp}{\displaystyle p\,(1-p)\,\ln\frac{1-p}{p}}=\ln\,\frac{ \displaystyle\ln\frac{1-p(0)}{p(0)}} {\displaystyle \ln\frac{1-p(t)}{p(t)}} \ ,
\end{equation}
or, equivalently,
\begin{subequations}\label{explicit}
\begin{eqnarray}\label{solution2}
p(t)&=& \frac{1}{\displaystyle 1+\left(\frac{1-p(0)}{p(0)}\right)^{\displaystyle \exp(-t/\tau)}}\\
&=& \frac{1}{2}+\frac{1}{2}\tanh\left(-\frac{1}{2}\exp(-t/\tau)\ln \frac{1-p(0)}{p(0)} \right) \ ,
\end{eqnarray}
\end{subequations}
from which we readily find $p(\infty)=1/2$ and
\begin{equation}
p(-\infty)= \frac{1}{\displaystyle 1+\left(\frac{1-p(0)}{p(0)}\right)^{\displaystyle \infty\vphantom{A^x}}}=\left\{ \begin{array}{cc}0&{\rm for}\ p(0)<1/2\\ 1&{\rm for}\ p(0)>1/2  \end{array}\right. \ .
\end{equation}

By analogy, the extension to this  degenerate case of the model
equation proposed in \cite{Lemanska} is $dp_j/dt=\upsilon(-\ln p_j
+ a_L)$ with $2a_L=\sum_{i=1}^2\ln p_i=\ln p+\ln (1-p)$ that is
(setting $\upsilon =1/2$ and adding, for clarity, the subscript
$L$)
\begin{equation}\label{twoleveldegenerateEquationLemanska}
\frac{dp_L }{dt}= \frac{1}{4}\ln\frac{1-p_L }{p_L } \ ,
\end{equation}
which yields the entropy generation rate
\begin{equation}\label{entropyratetwolevelL}\dot S_L = \frac{1}{4}\left(\ln\frac{1-p_L}{p_L}\right)^2 \ .\end{equation}

\begin{figure}[!h]
\begin{center}
\includegraphics[width=0.49\textwidth]{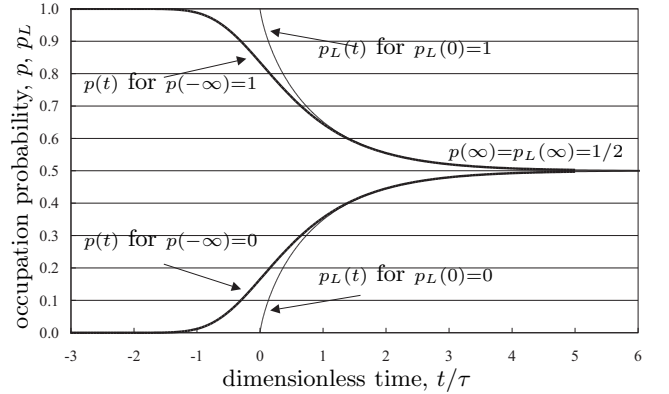}
\vskip5mm \caption{\label{FigureTwoLevelDegenerate}Comparison  of
the time dependences $p(t)$ and $p_L(t)$ respectively implied by
the rate equation (\ref{twoleveldegenerateEquation}) that we
propose and the rate equation
(\ref{twoleveldegenerateEquationLemanska}) discussed in [1].}
\end{center}
\end{figure}

Figure \ref{FigureTwoLevelDegenerate}  shows a comparison between
the time dependence (\ref{explicit}) implied by our rate equation
(\ref{twoleveldegenerateEquation}) and that obtained by numerical
solution  (by a standard Runge-Kutta method) of the rate equation
(\ref{twoleveldegenerateEquationLemanska}), for both $p(-\infty) =
0=p_L(0)$ and  $p(-\infty)= 1=p_L(0) $. For the purpose of
comparison, time $t=0$ is selected where $p_L(0)=0$ or 1 and the
initial state $p(0)$ is selected so as to emphasize that in the
limit as $t\rightarrow +\infty $ we have $p(t)\approx p_L(t)$. In
fact, we readily verify from both Eqs.
(\ref{twoleveldegenerateEquation}) and
(\ref{twoleveldegenerateEquationLemanska}) that the two time
dependences have the same asymptotic behavior as they approach the
equilibrium distribution, that is,
\begin{equation}
\frac{dp}{dt}\approx \frac{1}{2}-p \qquad {\rm and} \qquad \dot S\approx 4\,\left(\frac{1}{2}-p\right)^2\ .
\end{equation}

This simplest case brings out the evident different  behavior at
early times and the unphysical feature of the solution of Eq.\
(\ref{twoleveldegenerateEquationLemanska}) at $p_L=0$ where the
repopulation rate is infinite, implying that no unpopulated
eigenstate can survive unpopulated. Instead, our Eq.
(\ref{twoleveldegenerateEquation}) maintains unpopulated any
initially unpopulated eigenstate, and it also maintains relatively
little populated an initially little populated eigenstate for a
lapse of time that is quantified by Eq. (\ref{solution1}) and
depends on how close the initial value $p(0)$ is to zero. For
example, the time required to take $p(0)=10^{-2n}$ to
$p(t)=10^{-2}$ is $t\approx \ln n$ (that is, $t\approx \tau\ln
n$).

\begin{figure}[!h]
\begin{center}
\includegraphics[width=0.40\textwidth]{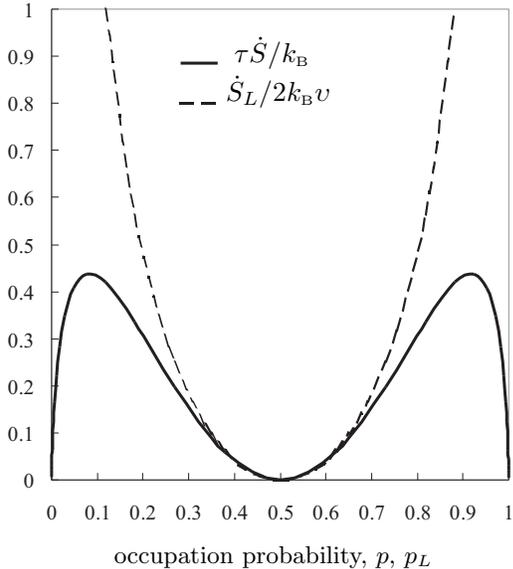}
\vskip5mm
\caption{\label{FigureEntropyTwoLevelDegenerate}Comparison
between the entropy generation rate $\dot S$ versus $p$ as given
by Eq.\ (\ref{entropyratetwolevel}) for $\Boltz=1$ and $\tau=1$
and $\dot S_L$ versus $p_L$ as given by Eq.\
(\ref{entropyratetwolevelL}) for $\Boltz=1$ and $\upsilon=1/2$.}
\end{center}
\end{figure}

Figure \ref{FigureEntropyTwoLevelDegenerate} shows a  plot of the
entropy generation rate $\dot S$ versus $p$ obtained from Eq.\
(\ref{entropyratetwolevel}) compared with $\dot S_L$ versus $p_L$
as obtained from Eq.\ (\ref{entropyratetwolevelL}), where again
the essential differences for small values of $p$ and $1-p$ are
singled out.

\section{Numerical results}\label{numerical}

The energy versus entropy diagram introduced by Gibbs  represents
the intersection with the $E$--$S$ plane of the
$E$--$S$--$V$--$\bn$ surface representing the stable thermodynamic
equilibrium states of a system, assuming that the energy
eigenvalues depend on the volume $V$ and the amounts of
constituents $\bn$, so that the surface is represented by the
so-called fundamental relation $S=S(E,\{e_j(V,\bn)\})$. In
\cite{Books} the use of such diagram has been extended to include
the projection onto the $E$--$S$ plane of all other states, i.e.,
not only the stable equilibrium states but also the
non-equilibrium and the non-stable equilibrium states, with given
fixed values of $V$ and $\bn$ and, therefore, a given fixed set of
energy eigenvalues. On such diagram, therefore, one point
represents in general a multitude of distributions, except at
every point of maximal entropy for each given value of $E$ ($V$
and $\bn$ are fixed) which corresponds to a unique canonical
distribution (\ref{canonical}), i.e., a unique stable
thermodynamic equilibrium state.

\begin{figure}[!h]
\begin{center}
\includegraphics[width=0.45\textwidth]{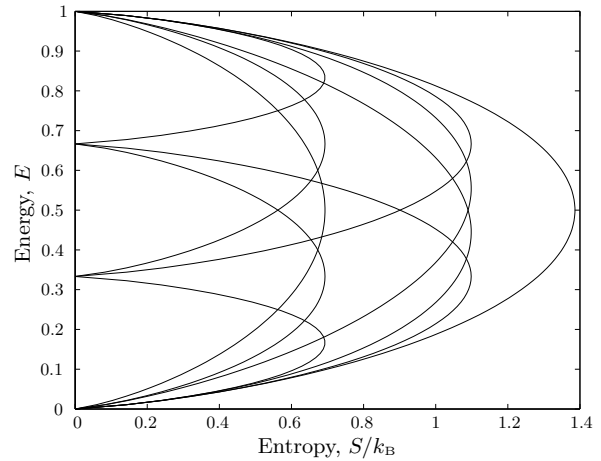}
\vskip5mm \caption{\label{DiagESwithPEstates}Representation  on an
energy versus entropy diagram (for $N=4$ and nondegenerate
eigenstates with energies $\be=[0,1/3,2/3,1]$) of the families of
possible canonical and partially-canonical equilibrium
distributions which in our dynamics are the only ones with zero
entropy generation rate. For example, a horizontal line at $E=0.4$
intersects seven different families of partially canonical
states.}
\end{center}
\end{figure}

For a four-level nondegenerate system, Figure
\ref{DiagESwithPEstates} represents on the diagram the families of
possible canonical (\ref{canonical}) and partially-canonical
(\ref{nondissipative}) equilibrium distributions which in our
dynamics are the only ones with zero entropy generation rate. We
recall that the slope of these curves is related to the parameter
$\beta^\pe(E,\d)$ because $\partial S^\pe(E,\d)/\partial E |_\d =
\Boltz\beta^\pe(E,\d)$, which for the canonical distribution (all
$\delta_i$'s equal to unity) is $\partial S(E)/\partial E=
\Boltz\beta(E)=1/T(E)$.

\begin{figure}[!h]
\begin{center}
\includegraphics[width=0.45\textwidth]{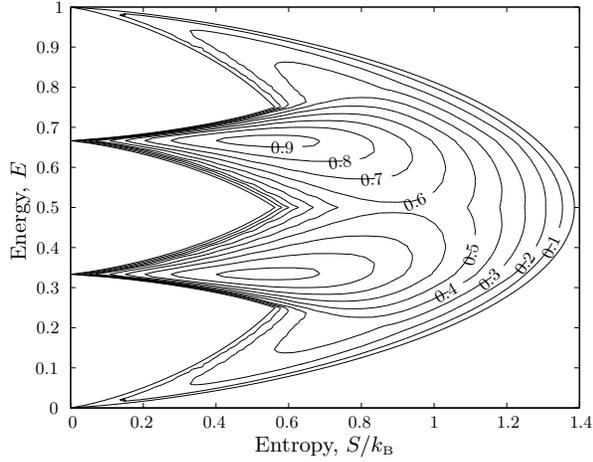}
\vskip5mm \caption{\label{isoDS3bnlabels}Representation on  an
energy versus entropy diagram (for $N=4$ and nondegenerate
eigenstates with energies $\be=[0,1/3,2/3,1]$) of the iso--$\dot
S_{\rm max}$ contour curves where $\dot S_{\rm max}$ represents at
each point in the diagram the highest value of the rate of entropy
generation $dS/dt$ according to Eq.\ (\ref{mainEquation}) among
all the possible distributions represented by that point.}
\end{center}
\end{figure}

The number of possible distributions that share a  given pair of
values of $E$ and $S$ is in general an $(N-3)$--fold infinity
except at maximal entropy for each value of $E$, where the
distribution is unique, and at few other notable exceptions such
as at minimal entropy for each given $E$ where the distribution
may be unique or sometimes many-fold. For all possible
distributions represented by a given point on the $E$--$S$
diagram, we may evaluate the rate of entropy generation $dS/dt$
according to Eq.\ (\ref{mainEquation}) and select the highest
value, that we denote by $\dot S_{\rm max}(E,S)$.  The result of
this numerical computation is sketched in Figure
\ref{isoDS3bnlabels} where the iso--$\dot S_{\rm max}$ contour
curves are plotted on the entire allowed domain on the energy
versus entropy diagram (of course, under the restriction to the
subset of states specified in the Introduction).

The next Figures show typical time dependences of  the occupation
probabilities that result from the numerical integration (by means
of a standard Runge-Kutta algorithm) of Eq.\ (\ref{mainEquation})
both in forward and backward time. All trajectories in these
Figures refer to a system with $N=4$ and nondegenerate eigenstates
with $\be=[0,1/3,2/3,1]$, and all have the same mean energy
$E=2/5$; they all tend, of course, to the canonical distribution
$\p^\se(2/5)=[0.3474,0.2722,0.2133,0.1671]$ that has inverse
temperature $\beta^\se(2/5)= 0.7321$. They are obtained by
assuming for all cases an initial distribution $\p(0)$ obtained by
perturbing the canonical distribution $\p^\se(E)$ [Eq.\
(\ref{canonical})] according to Eq.\ (\ref{renormalization}) with
the energy preserving perturbing factors defined as follows, for
$j=1,2,\dots,N$,
\begin{equation}\label{perturbation}
f_j=1-\lambda+\lambda\,\frac{p_j^\pe(E,\d)}{p_j^\se(E)}\quad{\rm with\ } 0<\lambda<1\ ,
\end{equation}
where $\lambda$ is otherwise arbitrary, and also $\d$  is
arbitrarily chosen among the possible vectors of 0's and 1's
compatible with the given value of $E$ and form
(\ref{nondissipative}) of the distribution $ \p^\pe(E,\d)$ (see
Figure \ref{DiagESwithPEstates}), where $\beta^\pe(E,\d)$ is
computed by solving the relation  $\sum_i \,p_i^\pe (E,\d) =E$.
For all subsequent Figures we use $\lambda=0.9$.

\begin{figure}[!h]
\begin{center}
\includegraphics[width=0.48\textwidth]{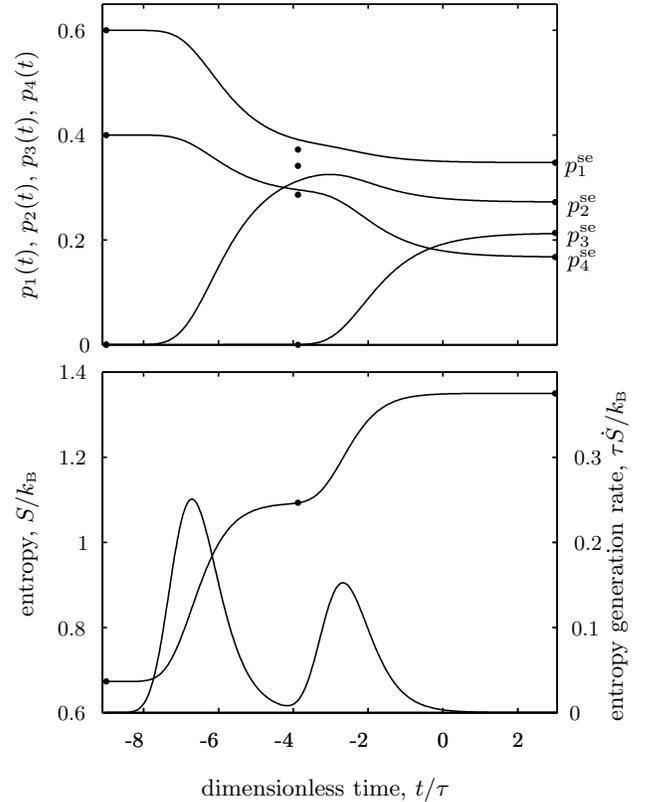}
\vskip5mm \caption{\label{FigureTrajectory&EntropyVsTime}Top:
typical time dependences of the occupation probabilities that
result from the numerical integration of Eq.\ (\ref{mainEquation})
both forward and backward in time, for $N=4$, $\be=[0,1/3,2/3,1]$,
energy $E=2/5$, initial state at $t=0$ from Eq.\
(\ref{perturbation2}) with $\lambda=0.9$ and $\d=[1,1,0,1]$. The
dots on the right represent the maximal entropy distribution; the
dots at the left represent the lowest-entropy or 'primordial'
distribution; the dots in the middle represent the $\p^\pe(E,\d)$
distribution used in Eq.\ (\ref{perturbation2}) to select the
$t=0$ state, plotted at the instant in time when the entropy of
the time-varying trajectory is equal to the entropy of the
$\p^\pe(E,\d)$ distribution.  Bottom: the corresponding time
dependence of the entropy (left axis) and the entropy generation
rate (right axis).}
\end{center}
\end{figure}

Figure \ref{FigureTrajectory&EntropyVsTime} shows the  time
dependence of the occupation probabilities that results under the
assumptions just cited using $E=2/5$, $\lambda=0.9$ and
$\d=[1,1,0,1]$ in Eq.\ (\ref{perturbation}) and subsequently
substituting in Eqs.\ (\ref{renormalization}), that is,
\begin{equation}\label{perturbation2}
\p(0)= \lambda \,\p^\pe(E,\d)+(1-\lambda)\,\p^\se(E) \ .
\end{equation}

It is noteworthy that when the trajectory gets  very close to the
partially-canonical unstable-equilibrium distribution
$\p^\pe(E=2/5,\d=[1,1,0,1])$ the entropy surface presents a local
'plateaux' and the entropy generation rate drops almost to zero,
but shortly after the trajectory bends in a direction of steeper
slope that drives the generation up again until the canonical
distribution $\p^\se(E)=[0.3474,0.2722,0.2133,0.1671]$ is finally
approached, with inverse temperature $\beta^\se(2/5)= 0.7321$. Of
course, the entropy is a monotonically increasing function of time
along the entire trajectory.

\begin{figure}[!h]
\begin{center}
\includegraphics[width=0.48\textwidth,height=1.1\textwidth]{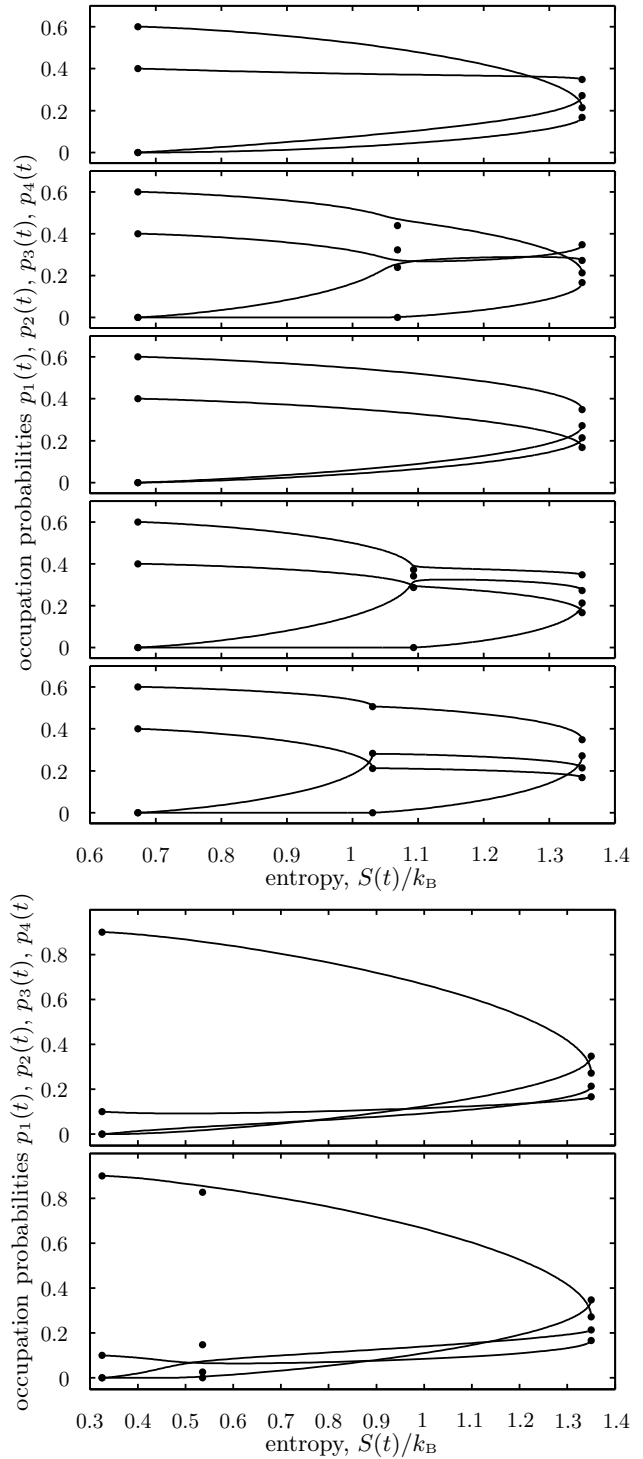}
\vskip5mm \caption{\label{FigureTrajectoriesVsEntropy}Plots  of
$p_i(t)$ versus $S(t)$ for seven sample time dependences of the
occupation probabilities that result from the numerical
integration of Eq.\ (\ref{mainEquation}) both forward and backward
in time, for different initial distributions. }
\end{center}
\end{figure}

\begin{figure}[!h]
\begin{center}
\includegraphics[width=0.48\textwidth]{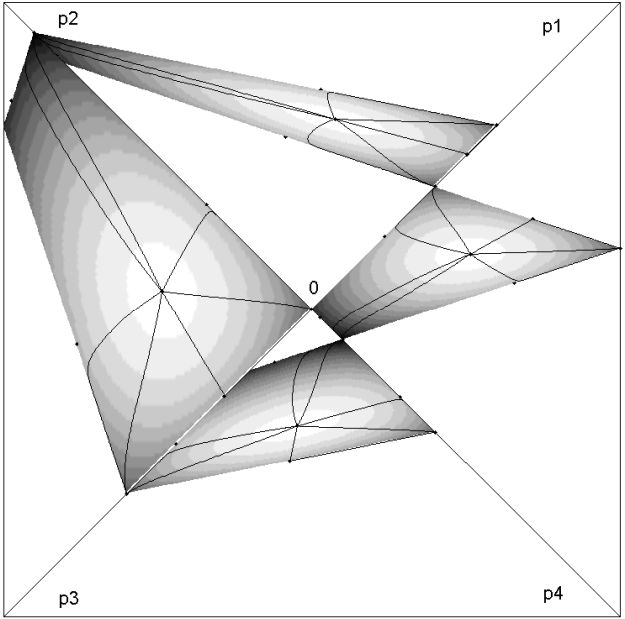}
\vskip5mm \caption{\label{FigureContourQuadrants}Each rotated
quadrant of the graph represents, for the corresponding pair of
occupation probabilities, a plot of the seven trajectories shown
in Figure \ref{FigureTrajectoriesVsEntropy} drawn over contour
plots of the entropy surface.}
\end{center}
\end{figure}

Figure \ref{FigureTrajectoriesVsEntropy} shows  the same
trajectory as well as six other trajectories, but instead of
plotting the time dependence of the occupation probabilities we
plot them against entropy. The initial (time $t=0$) distribution
used to obtain these seven sample trajectories are obtained from
Eq.\ (\ref{perturbation2}) with $E=2/5$, $\lambda=0.9$ and each of
the seven partially canonical states corresponding to the given
value of the energy. These seven states are easily identified on
the $E$--$S$ diagram in Figure \ref{DiagESwithPEstates} by drawing
a horizontal line at $E=0.4$. For the first, third, and sixth
trajectories we use the $\p^\pe(E,\d)$ states with $\d=[1,0,1,0]$,
$\d=[1,0,0,1]$ and $\d=[0,1,0,1]$, respectively, which [as
apparent from the subsequent Figure \ref{FigureContourQuadrants}]
are lowest-entropy boundary points of the entropy surface for the
given energy, and turn out to be also the 'primordial' states of
the corresponding trajectories. For the remaining trajectories we
use the $\p^\pe(E,\d)$ states with $\d=[1,1,1,0]$, $\d=[1,1,0,1]$,
$\d=[1,0,1,1]$, and $\d=[0,1,1,1]$, respectively. These too are
boundary points of the entropy surface, but they correspond to
partial maxima (over the subset of distributions with one
unoccupied eigenstate as specified by the corresponding zero
element of $\d$). It is seen that these partial maxima affect the
trajectories passing nearby by acting as partial attractors
especially in the initial phase of the time evolution.

Figure \ref{FigureContourQuadrants} is a more  elaborate
representation of the same seven trajectories. They are shown four
times from different perspectives on the backgorund of contour
plots of the entropy surface, for four pairs of occupation
probabilities. Indeed, for $N=4$ and fixed energy $E$, the number
of independent occupation probabilities is two. Thus for four
pairs of probabilities ($p_1$--$p_2$, $p_2$--$p_3$,$p_3$--$p_4$,
$p_4$--$p_1$), we draw the contour plot of the entropy surface
over the entire domain of allowed values (which of course are
contained in a triangular region of the first quadrant), and over
this plot we draw the seven trajectories (and the seven partially
canonical states used to choose them). To save space, we then
rotate each of the four graphs (respectively by 45, 135, 225, 315
degrees) and combine them on the same graph in Figure
\ref{FigureContourQuadrants}. The figure visualizes clearly that
the trajectories indeed follow paths of
locally-steepest-entropy-ascent and unfold smoothly also backward
in time to the 'primordial' states. We also note that these
lowest-entropy states exhibit a singular behavior in that, for
example, state [2/5,0,3/5,0] is the primordial state for two
entirely different trajectories, state [3/5,0,0,2/5] for three
other, and state [0,9/10,0,1/10] for the remaining two. Moreover,
the partially canonical states appear as partial attractors of
trajectories passing nearby, as seen quite clearly for the second,
fourth and fifth trajectory of Figure
\ref{FigureTrajectoriesVsEntropy}, which are partially attracted
by the partially canonical states with  $\d=[1,1,1,0]$,
$\d=[1,1,0,1]$ and $\d=[1,0,1,1]$, respectively.

\section{Conclusions}

The model we propose for the description of the  time evolution of
the occupation probabilities of a perturbed, isolated, physical
system with single-particle eigenstates with energies $e_i$ for
$i=$ 1, 2, \dots, $N$, is in good agreement with general
thermodynamic requirements such as energy conservation,
conservation of normalization and non-negativity of the
probabilities, entropy nondecrease, $E$-conditional stability of
the maximal-entropy canonical equilibrium states, $E$-conditional
non-stability of each non-maximal-entropy partially-canonical
equilibrium states, and existence and uniqueness of solutions for
all initial perturbed distributions, both in forward and backward
time. As in our previous work
\cite{Cimento1,Beretta,Attractor,thesis,Onsager}, the proposed
rate equations implement the fundamental ansatz that
nonequilibrium time dependence follows the path of
steepest-entropy-ascent (or, using the terminology adopted in
\cite{Gheorghiu2,Gheorghiu1}, maximal entropy generation).

The model can be readily generalized to include  additional
constraints and therefore adapted to other physical and
nonphysical (e.g., information theoretical, biological) problems
that obey the same maximal entropy formalism and the maximal
entropy generation rate ansatz. Using the formalism developed in
Section \ref{construction} it can even be readily generalized to
different entropy  functionals or nonlinear objective functionals
that may be relevant in many other contexts that share with the
present the basic mathematical framework.


\begin{thebibliography}{33}
\bibitem{Massieu}As we discuss in a forthcoming  paper to
be published elsewhere, the important relation
(\ref{fluctuationdissipation}) holds also for the quantum equation
proposed in \cite{thesis,Gheorghiu1,Cimento1}, i.e., also when the
restriction $[H,\rho]=0$, which defines the class of distributions
considered in the present paper, is removed.
\end{thebibliography}
\end{document}